# Sensor Artificial Intelligence and its Application to Space Systems – A White Paper

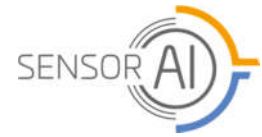

Anko Börner[1], Heinz-Wilhelm Hübers[1], Odej Kao[2], Florian Schmidt[2], Sören Becker[2], Joachim Denzler[3], Daniel Matolin[4], David Haber[5], Sergio Lucia[2], Wojciech Samek[6], Rudolph Triebel[7], Sascha Eichstädt[8], Felix Biessmann[9], Anna Kruspe[10], Peter Jung[2, 6], Manon Kok[11], Guillermo Gallego[2], Ralf Berger[1]

[1] German Aerospace Center, Institute of Optical Sensor Systems, Rutherfordstr. 2, 12489 Berlin Germany, anko.boerner@dlr.de

[2] Technische Universität Berlin, Marchstraße 23, 10587 Berlin, Germany

[3] Friedrich Schiller-Universität Jena, Friedrich-Schiller-Universität Jena, 07737 Jena, Germany

[4] PROPHESEE, 74 Rue du Faubourg Saint-Antoine, 75012 Paris, France

[5] Daedalean AG, Albisriederstrasse 199, 8047 Zürich, Switzerland

[6] Fraunhofer Heinrich-Hertz-Institut, Einsteinufer 37, 10587 Berlin, Germany

[7] German Aerospace Center, Institute of Robotics and Mechatronics, Münchener Straße 20, 82234 Weßling, Germany

[8] Physikalisch-Technische Bundesanstalt, Abbestraße 2, 10587 Berlin, Germany

[9] Beuth University of Applied Sciences Berlin, Luxemburger Str. 10, 13353 Berlin, Germany

[10] German Aerospace Center, Institute of Data Sciences, Mälzerstraße 3, 07745 Jena, Germany

[11] Delft University of Technology, Delft Center for Systems and Control, Mekelweg 2, 2628CD Delft, The Netherlands

## Introduction and Motivation

Information and communication technologies have accompanied our everyday life for years. A steadily increasing number of computers, cameras, mobile devices, etc. generate more and more data. Huge databases store this information and powerful networks distribute it on demand to almost any place on Earth. Based on complex processing algorithms and powerful hardware, research groups have explored new possibilities and developed solutions for challenging applications. Real-time face recognition on smartphones, decision support for medical diagnoses or driver assistance systems are impressive examples of what is possible today.

With a speed at least as fast as the technology evolves, expectations and requirements from potential users increase too. If applications such as the ones mentioned above should be massively deployed on safety critical environments, including cars, airplanes or hospitals, high technology readiness levels and outstanding quality guarantees are essential.

At the same time, it has become apparent that the amount and heterogeneity of the data captured by modern sensors cannot be completely analysed with known approaches. Classical algorithms relying on physical models are applicable if the problem can be well formulated in an analytical manner. For example, distances and areas can be measured in images if the object is depicted without occlusions and with sufficient contrast. Spectroscopic data can be evaluated if features are clearly detectable. Such tasks require calibration of the sensors as well as spatial and temporal referencing of data. Additionally, it is mostly assumed that the conditions under which the data were gathered (e.g. temperature or illumination) can be kept stable or at least monitored in laboratories or at assembly lines in factories.

Approaches dealing with more complex tasks, e.g. semantic classification or object detection in changing environments or operation "in the wild", lead to much higher error rates since there are no closed-form solutions – the parameter space is overwhelming large, there are too many unknowns and uncertainties. These approaches search for structure in the data, rather than for structure coming solely from a physical model. To deal with some of these challenges, great research efforts have been focused on the field of artificial intelligence (AI). AI has made enormous progress in multiple benchmarks in recent years, in particular machine learning (ML) has solved some challenges with remarkable results. Deep neural networks, recurrent neural networks and convolutional neural networks are standard technologies today in research for solving complex tasks in various fields such as computer vision, sensor fusion, communication or automatic decision making.

A major drawback of typical machine learning solutions is the black-box character of a neural network. While it might not be necessary to understand why a decision was made for an AI system to perform adequately, it is necessary to build trust in AI systems. As identified by the High-Level Expert Group on AI set up by the EU Commission [1], transparency, technical robustness and safety are key requirements that AI systems should meet to be deemed trustworthy. Transparency and explainability of AI methods and systems is one of the big current research topics [2]. An important progress has been done recently with different methods such as interpretable local surrogates, occlusion analysis or layer-wise relevant propagation.

To achieve technical robustness, scientist and engineers need to carefully design performance metrics that are relevant for the expected operating conditions of AI systems. Further important research topics contain the design of probabilistic performance guarantees and online monitoring systems that can make sure that the deployed systems are resilient and reliable and that reproducible results are obtained, including possible fall-back options. Still, trustworthy AI methods and software remain as an important research topic that needs to be further studied to allow its deployment on safety critical applications [3].

One example illustrating this melting pot of outstanding challenges, extraordinary requirements and unique possibilities is the field of astronautics which is currently undergoing a drastic change. AI can become a true key technology to enable a leap in quality in space research and space business.

## Sensor AI

With growing experience, both, the potential and the limitations of new AI methodologies are increasingly better understood. From our point of view, one essential element in the process chain is still missing – the sensor itself. With this white paper on 'Sensor AI' we would like to extend the focus of considerations to this missing element.

Typically, AI approaches start with data from which information and directions for action are derived. However, the processes and circumstances under which such data are collected and how they change over time or even over other domains (e.g. temperature or radiation) are rarely considered. The basic idea of the extended view is to incorporate sensor knowledge which is gained by modelling, characterization and application into data analysis. This holistic approach which considers entire signal chains from the origin to a data product will allow linking the information technology models to physical reality, to assign bits and bytes to SI units. A closer look at the sensors and their physical properties within AI approaches will lead to more robust and widely applicable algorithms.

Sensors are complex units that convert physical properties of the environment into (mostly) digital signals. Sensors and their working principles are well understood because scientists and engineers have been building detailed physical models during decades, and calibration of such sensors is well established in science and industry. Based on such detailed models, or digital twins, sensors can be characterized and monitored during their entire life cycle – from design over development to application.

This knowledge about sensors shall be used and combined with AI systems. Physical models and data based models as well as typically applied methods in these domains (e.g. controlling, optimizing and learning) shall be merged. This approach is called 'Sensor AI'. It is an interdisciplinary research field and involves several key technologies such as sensors, communication networks, processors, data sciences and artificial intelligence.

Within the field 'Sensor AI' the following research questions can be investigated (the list can and shall be extended):

1. How can physical models support data-based models and vice versa?
2. How to use data of AI systems to learn important insights about the sensor?
3. How can sensor knowledge be used to estimate uncertainties in AI systems?
4. Which existing computer hardware is able to run AI applications on embedded systems?
5. Which sensors are most suitable for AI applications?
6. How dedicated AI sensors can be developed?
7. How can sensor knowledge be used to learn in a more efficient way?
8. Which trade-offs are needed to run AI methods on existing computer hardware and how these trade-offs can be quantified?
9. How to certify AI systems?
10. How to distribute data processing considering sensor and network specific properties?
11. What is the best approach to compress, send or analyse large data with limited hardware and communication capabilities?
12. What is the relation between compression and interpretability and how can it be exploited to try to maximize both for a proposed system?
13. What is required to use AI methods and technologies also for applications and instruments in challenging environments, e.g. space?

## Sensor AI for space applications

While Sensor AI can benefit many different disciplines of science and engineering, we believe that optical spaceborne sensor systems, such as cameras or spectrometers, offer the perfect framework to showcase the potential of Sensor AI. Only a combined approach that considers hardware particularities, engineering knowledge and modern AI algorithms can exploit the full potential of a technology that needs to operate in a very demanding environment characterized by radiation, zero-gravity, vacuum as well as exceptional thermal and mechanical loads. In addition, resources for space instruments are extremely limited, e.g.:

- Electric power: typical processing units have a power consumption of a few 10W in average, powerful 500W GPU's are not feasible.
- Computer performance: Most of the CPUs flying currently are comparable to 80x86 technologies which are more than 20 years old.
- Data transfer: Data can be downlinked with a few hundred megabits per second if a ground station is available.

On the other side we face high requirements on crucial instrument parameters, e.g.:

- Data rates: 1GBit/s is standard today for cameras with high resolution in the spatial or spectral domain.
- Data quality: Todays space missions are unique, long-term and expensive projects with limited possibilities to update software in orbit. As a consequence, the entire development process of hardware and software is strictly standardized.
- Data availability: Data products shall be provided fast if not in real time.

Currently, "New Space" is re-shaping the entire field of space research and development as well as space utilization. After having navigation and communication operational in space for several decades already, launchers, Earth observation and even space transportation become commercial which opens up new possibilities but creates new challenges to engineers and scientists.

Based on these reflections, new directions of (research) activities can be identified and some of them are related to Sensor AI:

- Due to the discrepancy between the volume of acquired data and downlink capabilities there is an urgent need for a drastic data reduction (factor of 100 or more) which cannot be provided by ordinary compression approaches. In the near future, high level (maybe semantic) data processing on board satellites will be required. Nevertheless, any derived information must be assigned to qualifiable and certifiable metrics. We share the conviction that this will be doable only if physical sensor knowledge and modern AI technologies are taken into consideration.
- Spaceborne instruments can be built and operated in an optimal way only if they are considered as a part of a very complex system including the object of observation, the environment, the sensor itself and the data processing. Such complex systems cannot be modelled in a purely physical way, and so AI methods can help to approximate these models. This applies to the entire development cycle from design and optimization over characterization in laboratories to operation in space (e.g. for predictive control).

## Conclusion

Overall, astronautics is one of the most demanding frameworks one can imagine and of course, all these considerations can be transferred to other fields of application with similar challenging requirements like medical imaging or assistance systems in transportation.

The goal of this white paper is to establish 'Sensor AI' as a dedicated research topic. We want to exchange knowledge on the current state-of-the-art on Sensor AI, to identify synergies among research groups and thus boost the collaboration in this key technology for science and industry.

We expect that 'Sensor AI' will improve the quality of AI methods and systems, increase the efficiency of AI technologies and support embedded solutions. By this approach we aim "to bring AI into application" as stated as one emphasised mission in the Hightech strategy of the German government [4] where AI is acknowledged to be one of the key technologies for the next decade for all addressed relevant thematic fields. 'Sensor AI' is a highly relevant and dynamic topic with great potential. It will play a decisive role in autonomous driving as well as in areas of automated production and predictive maintenance or space research. It will contribute to minimizing risks for safety and to effective functioning of the liability regime, which was defined as a key problem in a white paper of the European Commission on Artificial Intelligence [1]. 'Sensor AI' is a promising technological niche being worth to focus on to strengthen the competitiveness of European industry and improve the wellbeing of citizens.